\documentclass[preprint,showpacs,preprintnumbers,amsmath,amssymb]{revtex4}

\usepackage{graphicx}
\usepackage{dcolumn}
\usepackage{bm}

\begin{document}


\title{Adjustable Microchip Ringtrap for Cold Atoms and Molecules}

\author{Paul M. Baker}
\email[]{AFRL.RVB.PA@hanscom.af.mil}
\affiliation{Air Force Research Laboratory, Hanscom AFB, MA 01731, USA}
\affiliation{Physics Department, Tufts University.}
 
\author{James A. Stickney}
\affiliation{Space Dynamics Laboratory, Bedford, MA 01730, USA}

\author{Matthew B. Squires}
\affiliation{Air Force Research Laboratory, Hanscom AFB, MA 01731, USA}

\author{James A. Scoville}
\affiliation{Air Force Research Laboratory, Hanscom AFB, MA 01731, USA}

\author{Evan J. Carlson}
\affiliation{Air Force Research Laboratory, Hanscom AFB, MA 01731, USA}

\author{Walter R. Buchwald}
\affiliation{Air Force Research Laboratory, Hanscom AFB, MA 01731, USA}  

\author{Steven M. Miller}
\affiliation{Air Force Research Laboratory, Hanscom AFB, MA 01731, USA}             

\begin{abstract}

We describe the design and function of a circular magnetic waveguide produced from wires on a microchip for atom interferometry using deBroglie waves. The guide is a two-dimensional magnetic minimum for trapping weak-field seeking states of atoms or molecules with a magnetic dipole moment. The design consists of seven circular wires sharing a common radius. We describe the design, the time-dependent currents of the wires and show that it is possible to form a circular waveguide with adjustable height and gradient while minimizing perturbation resulting from leads or wire crossings. This maximal area geometry is suited for rotation sensing with atom interferometry via the Sagnac effect using either cold atoms, molecules and Bose-condensed systems.

\end{abstract}


\date{\today}
\maketitle

\section{Introduction}

In recent years atom interferometry has been used to make precision measurements of various phenomena such as rotations, acceleration, gravity gradients and frequency separation of the hyperfine splitting for precision timekeeping \cite{Kasevich1997,Pritchard1997}. The sensitivity of these measurements are directly proportional to the interaction time or, in the case of rotation measurements, the area enclosed by the two separate interferometer paths. State of the art atom interferometers \cite{Kasevich1997,Pritchard1997,Prentiss2007, Shin2004, Wang2005, Sackett2006} use unconfined launched atom clouds with minimal external potentials during the interferometer cycle. Despite the success of unconfined atom interferometers there are limitations on the ultimate sensitivity which include: gravity accelerating the atoms \cite{Kasevich2000}, increasing the enclosed area requires increasing the size of the required magnetic shielding, longer interaction times leads to lower signal to noise resulting from lower densities because of ballistic expansion of the atomic cloud, limited dynamic range due to the atom clouds impacting the rotating vacuum enclosure during high dynamic rotations. One method of addressing these difficulties is to place the atoms in a confining potential. Several methods for building potentials suitable for use in confined atom interferometers have been developed \cite{Prentiss2007, Shin2004, Wang2005, Sackett2006, Nakagawa2006}.

One method for producing a potential suitable for confined atom interferometry involves fabricating small micrometer scale current carrying wires on an insulating substrate, commonly referred to as an atom chip. Current carrying wires on atom chips produce magnetic fields that can be used to trap atomic samples when prepared in a low-field seeking state. However, to be effective for trapped atom interferometry, the magnetic potential must be sufficiently uniform to avoid decoherence. The requirements on the smoothness of the potential are reduced if the separate atomic clouds propagate through reciprocal paths, canceling common mode noise \cite{Prentiss2007} and when the energy associated with the cloud is higher than the potential roughness \cite{Stickney2009}.

Atom interferometry for rotation sensing via the Sagnac effect is one of the most promising applications of trapped atom interferometers. To maximize the enclosed area, and thus sensitivity, the atomic clouds used in the interferometer should propagate in a circle. In this paper, we propose a method for fabricating an atom chip ring trap, specifically for use in atom interferometry. 

A challenge of the ring traps using atom chips is the elimination of potential imperfections resulting from the input leads. One method of avoiding the input leads is to use several turns in an effort to make the input lead perturbation small in comparison to the ring field \cite{Stamper-Kurn2005}. More recently magnetic induction has been proposed as means to avoid input leads \cite{Arnold2008}. In our previous paper \cite{Crookston2005} we proposed two sets of wires that provide two overlapping ring traps about a common radius and the ability to switch between the two in order to avoid the input leads. This method also provided a means of loading the atoms directly into the waveguide via a U-trap wire located adjacent to the ring to avoid atom losses \cite{Arnold2005}. An experimental limitation of this design was a fixed trapping height based upon the wire spacing and current ratios of the wires. Often the desired working distance is unknown and is not cost or time effective to redsesign and replace chips often. For this reason a chip design with adjustable trapping distance is strongly desired.

There are several reasons why a ring trap with an adjustable trapping height is experimentally useful. First, the lifetime of the atomic cloud trapped near the surface of an atom chip is limited by trap loss caused by Johnson noise photon induced spin flips \cite{Vuletic2004}. The number of Johnson noise photons produced is dependent on the temperature of the chip, which depends upon the current density of the wires and the thermal properties of the microchip. Because producing the same magnetic confinement further from a wire requires more current and therefore a higher chip temperature, optimizing the distance of the atoms from the chip is experimentally important. Also, the atom interferometer requires some form of splitting and re-combining of the atoms. A common method used is to apply an optical standing wave. Once again the distance of the atoms from the microchip is important because the Bragg scattering efficiency is reduced by the scattering of the laser beams off the chip surface. Finally, increasing the distance from the waveguide to the chip surface also averages small imperfection in the potential resulting from current fluctuations in the wires. For all of these reasons it is desirable to have a chip design that allows for an adjustable trapping distance by changing the wire currents.

\section{The 7-wire microchip ring trap design}

\begin{figure}
\begin{center}
\includegraphics[width = 9cm]{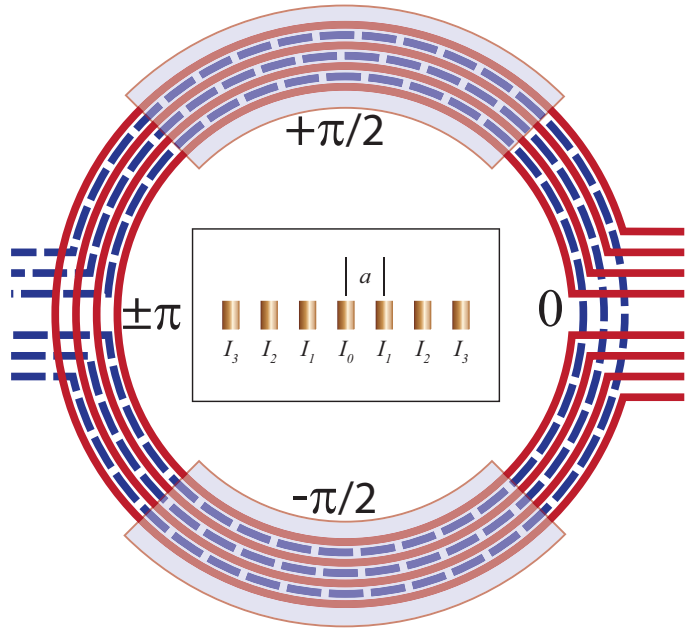}
\end{center}
\caption{(color online) 7-wire ringtrap layout including the appropriate current labels. The location of the wires used to form the 3-wire ring trap is shown as blue (dashed) lines with the input leads entering from the left and the wires used to produce the 4-wire ring trap are shown as red (solid) lines, with the input leads entering from the right. (Inset) The wire spacing is given by the parameter \textit{a} and the currents assigned to each wire are labeled $I_n$ accordingly.}
\label{fig:7WireLayout}
\end{figure}

Previous work has shown that a waveguide with a magnetic field minimum can be generated utilizing either 3 or 4 straight current carrying wires \cite{Cassettari2000, Thywissen1999}. Specific currents in the wires can be chosen, so as to produce a trapping potential some distance away from the wires.  In this paper it will be demonstrated that it is possible to use 7 concentric circular wires to produce a uniform ring trap waveguide that avoids perturbations resulting from the input leads. A schematic of our 7-wire ring trap is shown in Fig.~\ref{fig:7WireLayout}. Since there are a total of seven concentric current rings used to form this ring trap, we will refer to this geometry as a 7-wire ring trap for the remainder of this paper.  The primary advantage of using the 7-wire ring trap is that the distance of the ring trap from the atom chip can be varied simply by changing the currents in the wires.   

The operation of the ring trap is similar to our previous ring trap work \cite{Crookston2005}. Initially, the atoms are cooled below the recoil temperature and are loaded into the 3-wire waveguide at the position indicated by $0~\mbox{rad}$ in Fig. \ref{fig:7WireLayout}. The atoms are coherently split using a standing wave laser pulse \cite{Wang2005, Prentiss2007}, half of the atomic cloud is given a $2 \hbar k_l$ momentum kick clockwise and the other half is given a momentum kick counter-clockwise, where $k_l$ is the wave number of the lasers beams used to produce the standing wave. Since the atomic cloud is cooled below the recoil temperature the two clouds of atoms will spatially separate \cite{Sackett2006}.  Since the atoms are confined in the ring trap, the two atomic clouds will propagate in circular paths.  When the clouds have entered the regions located near $\pm \pi/2~ \mbox{rad}$, (shown as shaded boxes in Fig.~\ref{fig:7WireLayout}), currents in the 3-wire ring trap are slowly turned off, while the currents in the 4-wire ring trap are turned on. This switching prevents the atoms in the ring trap from experiencing perturbations in the potential due to the currents in the input leads. When the two clouds have entered the regions near $\pm \pi/2 ~\mbox{rad}$ for a second time, the currents are switched back into the 3-wire ring trap.  When the clouds return to their initial position, the are illuminated with a second standing wave pulse. By counting the number of atoms in the $0, \pm \hbar k_l$ momentum states, the Sagnac phase shift can be determined \cite{Sagnac1913, Post1967, Pritchard1997}.

In addition to the waveguide, a bias field can be applied to lift the minimum of the waveguide minimum from zero. The bias field can be created by either a central orthogonal current carrying wire or a Time-Orbiting potential (TOP) \cite{Cornell1995}. Since the atoms are cooled below the recoil temperature, a non-zero waveguide minimum is essential to reduce atom loss through Majorana spin flips. Although often experimentally necessary, the inclusion of a bias field is simple and its effects will be neglected for the remainder of this paper.

\section{Theoretical development}

Below we will introduce a simple theoretical model for a ring trap using $N$ (odd) concentric current carrying rings on the surface of an atom chip. The center ring has radius $R$ and the center to center distance between the rings is $a$. For simplicity, only the case where the radii of the rings, is much larger than the distance between them $R\gg a$, and much larger than the distance of the ring trap from the chip will be considered. Thus, we will neglect the effects due to the curvature of the wires. We will also treat the wires as thin and neglect any effects due to finite wire size. The lowest order effects due to wire curvature have been analyzed, but the resulting formula's provide little new insight into the operation of our ring trap.    

The vector potential due to a current carrying ring points in the azimuthal direction. In the limit of large radius $R$ the vector potential for $N$ (odd) equally spaced current carrying concentric rings is,

\begin{equation}
 A_\phi = - \frac{\mu_0}{4 \pi} \sum_{n=-(N-1)/2}^{(N-1)/2} I_n \ln \left[ (\delta r - n a)^2 + z^2 \right], \label{Nring1}
\end{equation}
where $I_n$ is the current in the $n$-th ring, $a$ is the spacing between the wires, $\delta r = r - R$ is the radial distance from the center ring to the field point, and $z$ is the height of the field point from the rings.   Expanding Eq.~(\ref{Nring1}) about the point $\delta r = 0$ and $z = z_0$ yields
\begin{eqnarray}
 A_\phi &=& -\frac{\mu_0}{4 \pi} \sum_n I_n \left[ -\frac{2 na}{(na)^2 + z_0^2} \delta r + \frac{2 z_0}{(na)^2 + z_0^2} \delta z \right. \nonumber \\
&+&
\left.
\frac{ \left( z_0 \delta r + na \delta z \right)^2 - \left(  z_0 \delta z - na \delta r \right)^2 }{ \left((na)^2 + z_0^2 \right)^2}
 \right],\label{Nring2}
\end{eqnarray}
where $\delta z = z - z_0$  and the constant terms have been dropped. From Eq.~(\ref{Nring2}) it is clear that the magnetic field is zero at the point $\delta r = 0$ and $z = z_0$ when the two linear terms in Eq.(~\ref{Nring2}) each vanish. The first term is zero when 
\begin{equation}
I_n = I_{-n}, \label{C1}
\end{equation} 
and the second term is zero when the currents are such that 
\begin{equation}
 0 = \sum_n \frac{I_n }{(n a)^2 + z_0^2}. \label{zeroLoc}
\end{equation}

When both Eqns.~(\ref{C1}) and (\ref{zeroLoc}) are fulfilled the vector potential Eq.~(\ref{Nring2}) becomes
\begin{eqnarray}
 A_\phi &=& \frac{\mu_0}{4 \pi} \sum_n I_n 
\frac{ z_0^2 - (na)^2  }{ \left( z_0^2 + (na)^2 \right)^2   }  
\left(\delta r^2  -  \delta z^2 \right).
\label{Nring3}
\end{eqnarray}
Taking the curl of Eq.~(\ref{Nring3}) yields the magnetic field components 
\begin{eqnarray}
 B_r = \frac{\mu_0}{2\pi} \sum_n I_n \frac{ z_0^2 - (na)^2  }{ \left( z_0^2 + (na)^2 \right)^2   }
\delta z    \nonumber \\
B_z = \frac{\mu_0}{2\pi} \sum_n I_n 
\frac{ z_0^2 - (na)^2  }{ \left( z_0^2 + (na)^2 \right)^2   }
\delta r
 .\label{BField}
\end{eqnarray}
Equations (\ref{BField}) show that the magnetic field near the minima is of the same form as a simple single wire wave guide \cite{Thywissen1999, Zimmermann2007}, with field gradient given by
\begin{equation}
  \label{eq:BGrad}
  B' = \frac{\mu_0}{2\pi} \sum_n I_n \frac{ z_0^2 - (na)^2  }{ \left( z_0^2 + (na)^2 \right)^2   }.
\end{equation}
Note that the sum still runs from $-(N-1)/2$ to $(N-1)/2$. 

We will now limit our discussion to the case of seven concentric rings, with seven independent currents. Equation (\ref{C1}) eliminates three of the currents, leaving us with four currents independent currents $I_0$, $I_1$, $I_2$ and $I_3$ (As shown in Fig.~\ref{fig:7WireLayout}). Initially, the atoms are loaded into a ring trap, \textit{i.e.} located at $0~\mbox{rad}$ below the center wire in Fig.~\ref{fig:7WireLayout}. To avoid the leads the currents in the wires of the 4-wire trap at this location must be zero, $I_1 = I_3 =0$. To satisfy Eq.~(\ref{zeroLoc}), the relation between the remaining to currents must be 
\begin{equation}
 I_2 = - I_0 \frac{4 a^2 + z_0^2}{2z_0^2}, \label{threeRatio}
\end{equation}
and the magnetic field gradient is
\begin{equation}
  \label{eq:Bp3Wire}
  B' = \frac{\mu_0 I_0}{2 \pi} \frac{8 a^2}{z_0^2(z_0^2 + 4 a^2)}.
\end{equation}

When the atomic clouds have propagated half way around the ring trap, they are near the $\pi~\mbox{rad}$ in Fig.~\ref{fig:7WireLayout}. To avoid the perturbations due to the leads of the 3-wire trap the currents at that position must vanish, $I_0 = I_2 = 0$. At this point the trap is formed only by the currents in the wires with current $I_1$ and $I_3$. To satisfy Eq.~(\ref{zeroLoc}), the relation between the nonzero currents must be
\begin{equation}
 I_3 = - I_1 \frac{9 a^2 + z_0^2}{(a^2 + z_0^2)}, \label{fourRatio}
\end{equation}
and the magnetic field gradient is
\begin{equation}
 B'  = \frac{\mu_0 I_1}{2 \pi} \frac{32 a^2 z^2_0 }{ (z_0^2+a^2)^2 (z_0^2 + 9 a^2) }. 
 \label{Bp4wire}
\end{equation}

When the atoms are in the region near $\pm \pi/2~\mbox{rad}$ as shown in Fig.~\ref{fig:7WireLayout}, there are no input leads and all seven wires can have nonzero current. In this situation Eq.~(\ref{zeroLoc}) has many solutions, but the simplest solution is to assume that both Eq.~(\ref{threeRatio}) and (\ref{fourRatio}) are fulfilled. There are now two free parameters to specify the magnetic field and the field gradient and can be expressed as
 
\begin{equation}
\label{eqn:Bp7Wire}
 B' = \frac{\mu_0}{2 \pi} \left( \frac{ 8 a^2 I_0 }{z_0^2 (z_0^2 + 4 a^2)} + \frac{32 a^2 z_0^2 I_1 }{ (z_0^2+a^2)^2 (z_0^2 + 9 a^2) } \right). 
\end{equation}

To avoid heating of the atomic gas as it moves around the ring, the magnetic field gradient $B'$ should be held constant. At time $t = 0$, $I_1 = 0$, and $I_0 = I_0(0)$. To hold the gradient constant, the time dependence of the current $I_1$ should be  

\begin{equation}
I_1(t) = \frac{(z_0^2 + a^2)^2 (z_0^2 + 9 a^2)}{4 z_0^4 (z_0^2 + 4 a^2)} \left( I_0(0) - I_0(t) \right).
\end{equation}

\section{Switching between the 3-wire and 4-wire ring trap}

\begin{figure}
\begin{center}
\includegraphics[width = 10cm]{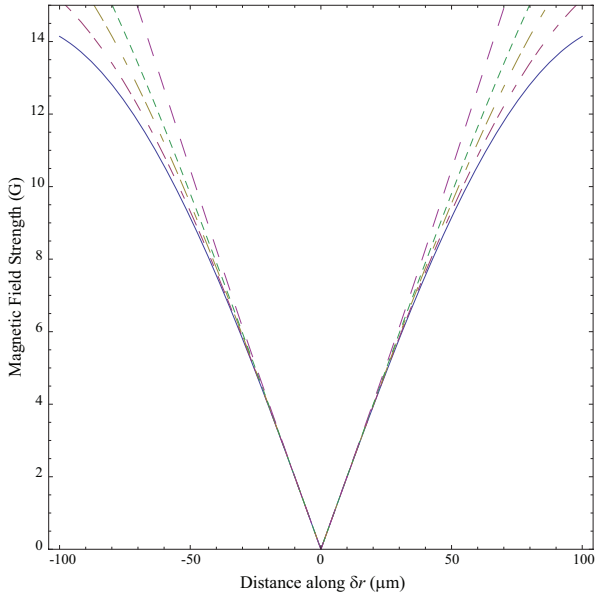}
\end{center}
\caption{(color online) Trapping potential in Gauss along $\delta r$-axis for $t=0-1$ in equal steps. Stepping through increments of $t$ is equivalent to turning off the current in the 3-wire waveguide and turning on the 4-wire waveguide. Notice that both the position of the minimum and the trapping shape near the minimum remain constant. $\delta r=0$ is the location of the center wire with current label $I_0$ and is perpendicular to the wires.}
\label{fig:XPlotTransfer}
\end{figure}

\begin{figure}
\begin{center}
\includegraphics[width = 10cm]{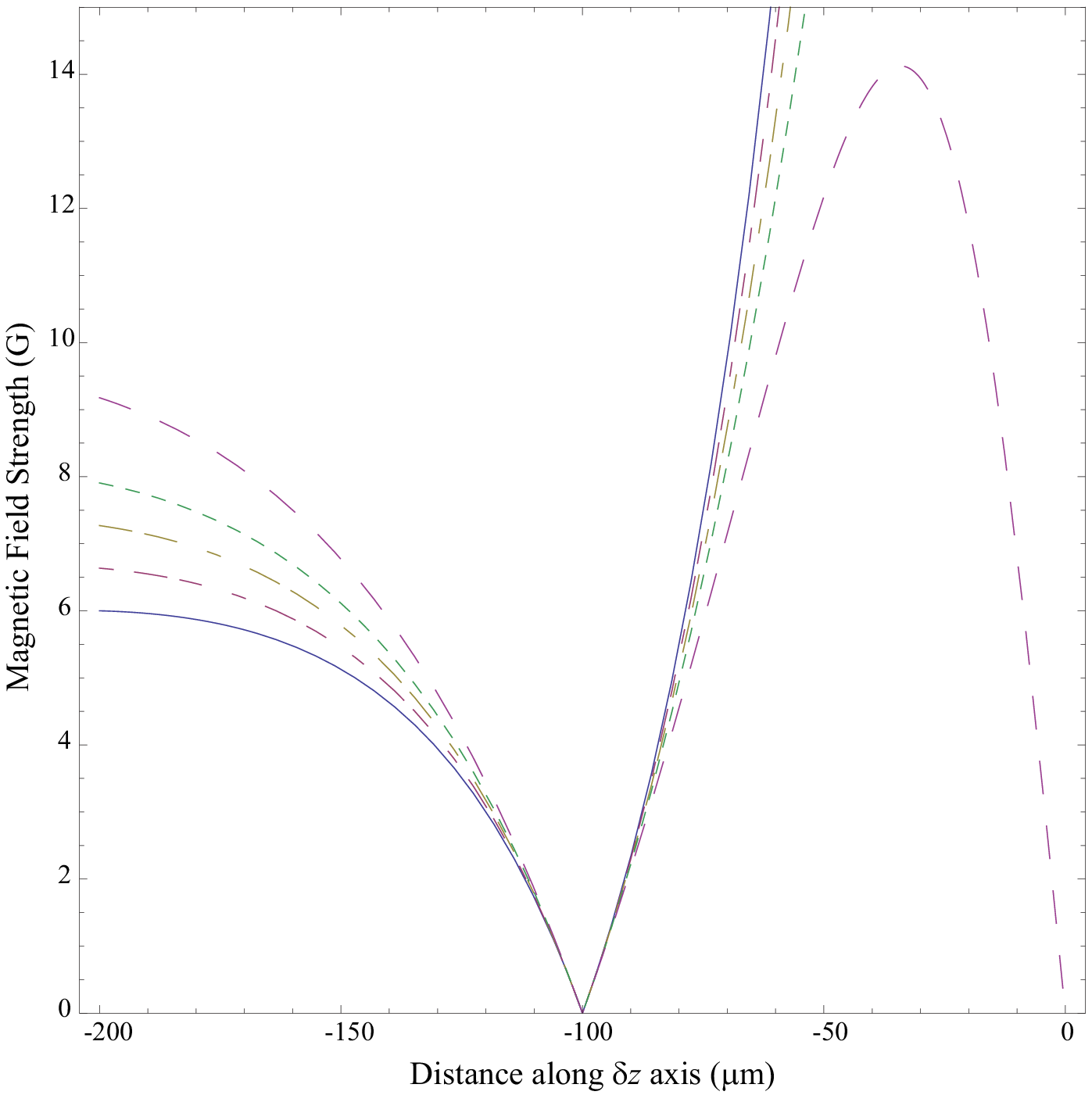}
\end{center}
\caption{(color online) Trapping potential below the wires is given in Gauss along $\delta z$-axis for $t=0-1$ in equal steps. Stepping through increments of $t$ is equivalent to turning off the current in the 3-wire waveguide and turning on the 4-wire waveguide. Notice that both the position of the minimum and the trapping shape near the minimum remain constant.}
\label{fig:ZPlotTransfer}
\end{figure}

To demonstrate the uniformity of the trapping potential while transferring from the 3-wire guide to the 4-wire guide, plots are given in Figs.~\ref{fig:XPlotTransfer} and \ref{fig:ZPlotTransfer} of the magnitude of magnetic field strength in steps of \textit{t}. In Figs.~\ref{fig:XPlotTransfer} and \ref{fig:ZPlotTransfer} , $a = 50~\mu \mbox{m}$, $z_0 = 100~\mu \mbox{m}$, and the current $I_{ref}=0.5~\mbox{A}$ was chosen to give a gradient $B' = 1000~\frac{\mbox{G}}{\mbox{cm}}$. Time dependent currents are given as follows:
\begin{equation}
I_0(t) = I_{ref}\frac{(t_{max}-t)}{t_{max}}
\end{equation}
\begin{equation}
I_1(t) = I_{ref} \frac{(z_0^2 + a^2)^2 (z_0^2 + 9 a^2)}{4 z_0^4 (z_0^2 + 4 a^2)}\frac{t}{t_{max}},
\end{equation}
where $t_{max}=1.0$ and \textit{t} was chosen to allow each waveguide to have values of $I_{ref}$ between 0 and 1. This procedure serves to switch between the 3-wire and the 4-wire waveguide in a linear manner. Notice that both the location of the minimum and the shape of the potential near the minimum remain constant as seen in Figs.~\ref{fig:XPlotTransfer} and~\ref{fig:ZPlotTransfer}.

\begin{figure}
\begin{center}
\includegraphics[width=9cm]{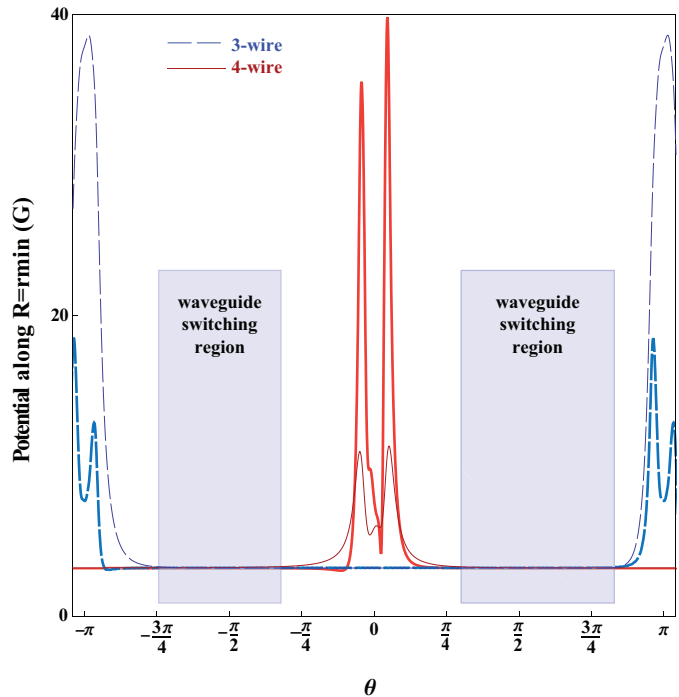}
\end{center}
\caption{Maintaining a constant waveguide minimum during the transfer from 3-wire to 4-wire waveguide is represented by the uniform flatness of the potential in the switching areas. Solutions are given for arcs connected to wires representing the input leads. The blue (dashed) curve is the magnitude of the magnetic field when only the three-wire ring trap has nonzero current and the red (solid) curve is the field when only the four-wire ring trap has nonzero current. There are two solutions depending upon the symmetry of the current direction, only the symmetric current configuration is discussed in this paper, however the anti-symmetric case is shown above with thicker lines for completeness and to illustrate experimental flexibility. A small constant bias field is applied to lift the minimum from zero as would be required experimentally to reduce Majorana losses.}
\label{fig:7WireOverlap}
\end{figure}

To characterize the effects of the inputs lead and curvature of the wires, we numerically calculated the magnetic field strength along a constant radius as shown in Fig.~\ref{fig:7WireOverlap}. In this calculation the current carrying wires are assumed to be thin, which is valid when the distance of the ring trap from the wires is larger than the size of the wires. Since the wires cannot be larger than the spacing between them, the thin wire approximation is always valid when $z_0 \gg a$.

We have performed numerical and analytic studies of the effects on the ring trap due to the finite curvature of the wires used to form the ring trap. Our results show that the curvature of the wires causes a small shift in the location of the ring trap towards the center; the ring trap is no longer located directly above the center wire. This shift can be corrected by making a correction to Eq.~(\ref{C1}) on the order of $na/R$. A more complete discussion of the curvature effects will be presented in future work. There are also corrections to Eqns. (\ref{threeRatio}), (\ref{fourRatio}), and (\ref{eqn:Bp7Wire}) on the order of $na/R$. The leads add yet another perturbation that shifts the location of the minimum and alters the shape of the potential. This effect is small in the region of interest and the approach of this paper is to switch the input leads before this perturbation is significant. None of these corrections have an effect on the ability to smoothly guide atom clouds around a ring.

In Fig. \ref{fig:7WireOverlap} the magnitude of the magnetic field is plotted along $\theta$ at a fixed radius at the field minimum for the same values given above. As previously mentioned the minimum will deviate slightly from $r_{min}$ as $\theta$ approaches the leads however the current in the leads is being turned down as the atoms approach reducing this perturbation. When the atomic clouds are located between $\theta = -\pi/4$ and $\theta = \pi/4$, the magnetic field has large perturbations due to the four wire ring trap's leads. Similarly, when the atomic cloud is located between $\theta = 3 \pi/4$ and $\theta = -3 \pi / 4$ the magnetic field has large perturbations due to the leads of the three wire ring trap. However, the atomic clouds are between $\theta = \pm \pi/4$ and $\theta = \pm 3 \pi / 4$, there are no perturbations due to either the three or four wire leads. This numerical solution demonstrates that there is a large region where the current can be switched between the two sets of wires.  

\section{Conclusions}

We have designed and developed a 7-wire ring trap with adjustable height that encloses area and avoids perturbation from input leads. We have introduced a 1-D theoretical model demonstrating it is possible to fabricate a chip where the currents can be switched between the three and four wire ring trap while holding the minimums location and gradient constant. We have numerically analyzed the effects of the input leads and shown that there is a large switching region where the perturbations due to the input leads of both the three and four wire rings can be avoided. Finally, we have briefly discussed our preliminary results of the effects of the curvature of the wires on the ring trap.  


The choice of atomic cloud temperature plays a pivotal role in the ring trap operation. Bose Einstein Condensates in microchip waveguides can suffer from fragmentation and de-phasing which are undesirable in atom interferometers. Recently, Bouchoule \textit{et. al.}, \cite{Bouchoule2007a, Bouchoule2007} has demonstrated a possible solution to the fragmentation issue and it is possible to operate the 7-wire ring trap in a manner that makes use of this technique. A BEC can also have reduced coherence times resulting from potential noise and mean field interactions. The short coherence times resulting from mean field interactions are density dependent \cite{Nakagawa2006, Sackett2006}; therefore a tightly confined BEC would have additional dispersion and dephasing. The 7-wire ring trap design has the additional feature of an adjustable gradient and by adjusting the gradient and utilizing dilute samples the 7-wire ring trap would be able to reduce the atom-atom interactions. Furthermore a weaker transverse confinement allows for more transverse oscillation which can be used for dispersion management \cite{Stamper-Kurn2006}. The ability to adjust the gradient thus affords more experimental flexibility.

It is experimentally useful to adjust the ring trap radius, \textit{i.e} for a choice of interferometer interrogation time. De-coupling of temporal and spatial sources of error is useful for systematically identifying and eliminating sources of noise and atom loss. Also, the possibility of adjusting the radius of the waveguide dynamically allows for the study of the coupling of longitudinal and transverse modes that could be used to damp out transverse oscillations if desired and help overlap the clouds at the recombination point \cite{Stamper-Kurn2005,Stamper-Kurn2006}. This concept can be extended to an \textit{N}-wire ring trap where the radial location of the minimum can be adjusted.

Finally it should be remarked that care must be taken during loading of the ring trap. Small shot-to-shot uncertainty in the initial momentum of the atom cloud, resulting from poor loading or coupling is sufficient to mask the small phase shifts resulting from rotation \cite{Sackett2009}. The 7-wire ring trap and ring traps in general may require additional loading wires that allow the atom cloud to come to equilibrium before optical splitting.  

\section{Acknowledgments}

The authors acknowledge support from the Air Force Office of Scientific Research under program/task 2301DS/03VS02COR and DARPA gBECi program.

\bibliography{ref}

\begin{thebibliography}{24}
\expandafter\ifx\csname natexlab\endcsname\relax\def\natexlab#1{#1}\fi
\expandafter\ifx\csname bibnamefont\endcsname\relax
  \def\bibnamefont#1{#1}\fi
\expandafter\ifx\csname bibfnamefont\endcsname\relax
  \def\bibfnamefont#1{#1}\fi
\expandafter\ifx\csname citenamefont\endcsname\relax
  \def\citenamefont#1{#1}\fi
\expandafter\ifx\csname url\endcsname\relax
  \def\url#1{\texttt{#1}}\fi
\expandafter\ifx\csname urlprefix\endcsname\relax\def\urlprefix{URL }\fi
\providecommand{\bibinfo}[2]{#2}
\providecommand{\eprint}[2][]{\url{#2}}

\bibitem[{\citenamefont{Gustavson et~al.}(1997)\citenamefont{Gustavson, Bouyer,
  and Kasevich}}]{Kasevich1997}
\bibinfo{author}{\bibfnamefont{T.~L.} \bibnamefont{Gustavson}},
  \bibinfo{author}{\bibfnamefont{P.}~\bibnamefont{Bouyer}}, \bibnamefont{and}
  \bibinfo{author}{\bibfnamefont{M.~A.} \bibnamefont{Kasevich}},
  \bibinfo{journal}{Phys. Rev. Lett.} \textbf{\bibinfo{volume}{78}},
  \bibinfo{pages}{2046} (\bibinfo{year}{1997}).

\bibitem[{\citenamefont{Lenef et~al.}(1997)\citenamefont{Lenef, Hammond, Smith,
  Chapman, Rubenstein, and Pritchard}}]{Pritchard1997}
\bibinfo{author}{\bibfnamefont{A.}~\bibnamefont{Lenef}},
  \bibinfo{author}{\bibfnamefont{T.~D.} \bibnamefont{Hammond}},
  \bibinfo{author}{\bibfnamefont{E.~T.} \bibnamefont{Smith}},
  \bibinfo{author}{\bibfnamefont{M.~S.} \bibnamefont{Chapman}},
  \bibinfo{author}{\bibfnamefont{R.~A.} \bibnamefont{Rubenstein}},
  \bibnamefont{and} \bibinfo{author}{\bibfnamefont{D.~E.}
  \bibnamefont{Pritchard}}, \bibinfo{journal}{Phys. Rev. Lett.}
  \textbf{\bibinfo{volume}{78}}, \bibinfo{pages}{760} (\bibinfo{year}{1997}).

\bibitem[{\citenamefont{Wu et~al.}(2007)\citenamefont{Wu, E., and
  Prentiss}}]{Prentiss2007}
\bibinfo{author}{\bibfnamefont{S.}~\bibnamefont{Wu}},
  \bibinfo{author}{\bibfnamefont{S.}~\bibnamefont{E.}}, \bibnamefont{and}
  \bibinfo{author}{\bibfnamefont{M.}~\bibnamefont{Prentiss}},
  \bibinfo{journal}{Phys. Rev. Lett.} \textbf{\bibinfo{volume}{99}},
  \bibinfo{pages}{173201} (\bibinfo{year}{2007}).

\bibitem[{\citenamefont{Shin et~al.}(2004)\citenamefont{Shin, Pasquini,
  Pritchard, and Leanhardt}}]{Shin2004}
\bibinfo{author}{\bibfnamefont{Y.}~\bibnamefont{Shin}},
  \bibinfo{author}{\bibfnamefont{T.~A.} \bibnamefont{Pasquini}},
  \bibinfo{author}{\bibfnamefont{D.~E.} \bibnamefont{Pritchard}},
  \bibnamefont{and} \bibinfo{author}{\bibfnamefont{A.~E.}
  \bibnamefont{Leanhardt}}, \bibinfo{journal}{Phys. Rev. Lett.}
  \textbf{\bibinfo{volume}{92}}, \bibinfo{pages}{050405}
  (\bibinfo{year}{2004}).

\bibitem[{\citenamefont{Wang et~al.}(2005)\citenamefont{Wang, Anderson, Bright,
  Cornell, Diot, Kishimoto, Prentiss, Saravanan, Segal, and Wu}}]{Wang2005}
\bibinfo{author}{\bibfnamefont{Y.}~\bibnamefont{Wang}},
  \bibinfo{author}{\bibfnamefont{D.}~\bibnamefont{Anderson}},
  \bibinfo{author}{\bibfnamefont{V.~M.} \bibnamefont{Bright}},
  \bibinfo{author}{\bibfnamefont{E.~A.} \bibnamefont{Cornell}},
  \bibinfo{author}{\bibfnamefont{Q.}~\bibnamefont{Diot}},
  \bibinfo{author}{\bibfnamefont{T.}~\bibnamefont{Kishimoto}},
  \bibinfo{author}{\bibfnamefont{M.}~\bibnamefont{Prentiss}},
  \bibinfo{author}{\bibfnamefont{R.~A.} \bibnamefont{Saravanan}},
  \bibinfo{author}{\bibfnamefont{S.~R.} \bibnamefont{Segal}}, \bibnamefont{and}
  \bibinfo{author}{\bibfnamefont{S.}~\bibnamefont{Wu}}, \bibinfo{journal}{Phys.
  Rev. Lett.} \textbf{\bibinfo{volume}{94}}, \bibinfo{pages}{090405}
  (\bibinfo{year}{2005}).

\bibitem[{\citenamefont{Garcia et~al.}(2006)\citenamefont{Garcia, Deissler,
  Hughes, Reeves, and Sackett}}]{Sackett2006}
\bibinfo{author}{\bibfnamefont{O.}~\bibnamefont{Garcia}},
  \bibinfo{author}{\bibfnamefont{B.}~\bibnamefont{Deissler}},
  \bibinfo{author}{\bibfnamefont{K.~J.} \bibnamefont{Hughes}},
  \bibinfo{author}{\bibfnamefont{J.~M.} \bibnamefont{Reeves}},
  \bibnamefont{and} \bibinfo{author}{\bibfnamefont{C.~A.}
  \bibnamefont{Sackett}}, \bibinfo{journal}{Phys. Rev. A}
  \textbf{\bibinfo{volume}{74}}, \bibinfo{pages}{031601(R)}
  (\bibinfo{year}{2006}).

\bibitem[{\citenamefont{Gustavson et~al.}(2000)\citenamefont{Gustavson,
  Landragin, and Kasevich}}]{Kasevich2000}
\bibinfo{author}{\bibfnamefont{T.~L.} \bibnamefont{Gustavson}},
  \bibinfo{author}{\bibfnamefont{A.}~\bibnamefont{Landragin}},
  \bibnamefont{and} \bibinfo{author}{\bibfnamefont{M.~A.}
  \bibnamefont{Kasevich}}, \bibinfo{journal}{Class. Quant. Grav.}
  \textbf{\bibinfo{volume}{17}}, \bibinfo{pages}{2385} (\bibinfo{year}{2000}).

\bibitem[{\citenamefont{Horikoshi and Nakagawa}(2006)}]{Nakagawa2006}
\bibinfo{author}{\bibfnamefont{M.}~\bibnamefont{Horikoshi}} \bibnamefont{and}
  \bibinfo{author}{\bibfnamefont{K.}~\bibnamefont{Nakagawa}},
  \bibinfo{journal}{Phys. Rev. A} \textbf{\bibinfo{volume}{74}},
  \bibinfo{pages}{031602(R)} (\bibinfo{year}{2006}).

\bibitem[{\citenamefont{Stickney et~al.}(2009)\citenamefont{Stickney, Squires,
  Scoville, Baker, and Miller}}]{Stickney2009}
\bibinfo{author}{\bibfnamefont{J.~A.} \bibnamefont{Stickney}},
  \bibinfo{author}{\bibfnamefont{M.~B.} \bibnamefont{Squires}},
  \bibinfo{author}{\bibfnamefont{J.}~\bibnamefont{Scoville}},
  \bibinfo{author}{\bibfnamefont{P.}~\bibnamefont{Baker}}, \bibnamefont{and}
  \bibinfo{author}{\bibfnamefont{S.}~\bibnamefont{Miller}},
  \bibinfo{journal}{Phys. Rev. A} \textbf{\bibinfo{volume}{79}},
  \bibinfo{pages}{013618} (\bibinfo{year}{2009}).

\bibitem[{\citenamefont{Gupta et~al.}(2005)\citenamefont{Gupta, Murch, Moore,
  Purdy, and Stamper-Kurn}}]{Stamper-Kurn2005}
\bibinfo{author}{\bibfnamefont{S.}~\bibnamefont{Gupta}},
  \bibinfo{author}{\bibfnamefont{K.~W.} \bibnamefont{Murch}},
  \bibinfo{author}{\bibfnamefont{K.~L.} \bibnamefont{Moore}},
  \bibinfo{author}{\bibfnamefont{T.~P.} \bibnamefont{Purdy}}, \bibnamefont{and}
  \bibinfo{author}{\bibfnamefont{D.~M.} \bibnamefont{Stamper-Kurn}},
  \bibinfo{journal}{Phys. Rev. Lett.} \textbf{\bibinfo{volume}{95}},
  \bibinfo{pages}{143201} (\bibinfo{year}{2005}).

\bibitem[{\citenamefont{Griffin et~al.}(2008)\citenamefont{Griffin, Riis, and
  Arnold}}]{Arnold2008}
\bibinfo{author}{\bibfnamefont{P.~F.} \bibnamefont{Griffin}},
  \bibinfo{author}{\bibfnamefont{E.}~\bibnamefont{Riis}}, \bibnamefont{and}
  \bibinfo{author}{\bibfnamefont{A.~S.} \bibnamefont{Arnold}},
  \bibinfo{journal}{Phys. Rev. A} \textbf{\bibinfo{volume}{77}},
  \bibinfo{pages}{051402(R)} (\bibinfo{year}{2008}).

\bibitem[{\citenamefont{Crookston et~al.}(2005)\citenamefont{Crookston, Baker,
  and Robinson}}]{Crookston2005}
\bibinfo{author}{\bibfnamefont{M.~B.} \bibnamefont{Crookston}},
  \bibinfo{author}{\bibfnamefont{P.~M.} \bibnamefont{Baker}}, \bibnamefont{and}
  \bibinfo{author}{\bibfnamefont{M.~P.} \bibnamefont{Robinson}},
  \bibinfo{journal}{J. Phys. B} \textbf{\bibinfo{volume}{38}},
  \bibinfo{pages}{3289} (\bibinfo{year}{2005}).

\bibitem[{\citenamefont{Arnold et~al.}(2006)\citenamefont{Arnold, Garvie, and
  Riis}}]{Arnold2005}
\bibinfo{author}{\bibfnamefont{A.~S.} \bibnamefont{Arnold}},
  \bibinfo{author}{\bibfnamefont{C.~S.} \bibnamefont{Garvie}},
  \bibnamefont{and} \bibinfo{author}{\bibfnamefont{E.}~\bibnamefont{Riis}},
  \bibinfo{journal}{Phys. Rev. A} \textbf{\bibinfo{volume}{73}},
  \bibinfo{pages}{041606(R)} (\bibinfo{year}{2006}).

\bibitem[{\citenamefont{Lin et~al.}(2004)\citenamefont{Lin, Teper, Chin, and
  Vuleti\'{c}}}]{Vuletic2004}
\bibinfo{author}{\bibfnamefont{Y.}~\bibnamefont{Lin}},
  \bibinfo{author}{\bibfnamefont{I.}~\bibnamefont{Teper}},
  \bibinfo{author}{\bibfnamefont{C.}~\bibnamefont{Chin}}, \bibnamefont{and}
  \bibinfo{author}{\bibfnamefont{V.}~\bibnamefont{Vuleti\'{c}}},
  \bibinfo{journal}{Phys. Rev. Lett.} \textbf{\bibinfo{volume}{92}},
  \bibinfo{pages}{050404} (\bibinfo{year}{2004}).

\bibitem[{\citenamefont{Cassettari et~al.}(2000)\citenamefont{Cassettari,
  Hessmo, Folman, Maier, and Schmiedmayer}}]{Cassettari2000}
\bibinfo{author}{\bibfnamefont{D.}~\bibnamefont{Cassettari}},
  \bibinfo{author}{\bibfnamefont{B.}~\bibnamefont{Hessmo}},
  \bibinfo{author}{\bibfnamefont{R.}~\bibnamefont{Folman}},
  \bibinfo{author}{\bibfnamefont{T.}~\bibnamefont{Maier}}, \bibnamefont{and}
  \bibinfo{author}{\bibfnamefont{J.}~\bibnamefont{Schmiedmayer}},
  \bibinfo{journal}{Phys. Rev. Lett.} \textbf{\bibinfo{volume}{85}},
  \bibinfo{pages}{5483} (\bibinfo{year}{2000}).

\bibitem[{\citenamefont{Thywissen et~al.}(1999)\citenamefont{Thywissen,
  Olshanii, Drndi\'{c}, Westervelt, and Prentiss}}]{Thywissen1999}
\bibinfo{author}{\bibfnamefont{J.~H.} \bibnamefont{Thywissen}},
  \bibinfo{author}{\bibfnamefont{M.}~\bibnamefont{Olshanii}},
  \bibinfo{author}{\bibfnamefont{M.}~\bibnamefont{Drndi\'{c}}},
  \bibinfo{author}{\bibfnamefont{R.~M.} \bibnamefont{Westervelt}},
  \bibnamefont{and} \bibinfo{author}{\bibfnamefont{M.}~\bibnamefont{Prentiss}},
  \bibinfo{journal}{Eur. Phys. D} \textbf{\bibinfo{volume}{7}},
  \bibinfo{pages}{361} (\bibinfo{year}{1999}).

\bibitem[{\citenamefont{Sagnac}(1913)}]{Sagnac1913}
\bibinfo{author}{\bibfnamefont{G.}~\bibnamefont{Sagnac}}, \bibinfo{journal}{C.
  R. Acad Sci.} \textbf{\bibinfo{volume}{95}}, \bibinfo{pages}{708}
  (\bibinfo{year}{1913}).

\bibitem[{\citenamefont{Post}(1967)}]{Post1967}
\bibinfo{author}{\bibfnamefont{E.~J.} \bibnamefont{Post}},
  \bibinfo{journal}{Rev. Mod. Phys.} \textbf{\bibinfo{volume}{39}},
  \bibinfo{pages}{475} (\bibinfo{year}{1967}).

\bibitem[{\citenamefont{Petrich et~al.}(1995)\citenamefont{Petrich, Anderson,
  Ensher, and Cornell}}]{Cornell1995}
\bibinfo{author}{\bibfnamefont{W.}~\bibnamefont{Petrich}},
  \bibinfo{author}{\bibfnamefont{M.~H.} \bibnamefont{Anderson}},
  \bibinfo{author}{\bibfnamefont{J.~R.} \bibnamefont{Ensher}},
  \bibnamefont{and} \bibinfo{author}{\bibfnamefont{E.~A.}
  \bibnamefont{Cornell}}, \bibinfo{journal}{Phys. Rev. Lett.}
  \textbf{\bibinfo{volume}{74}}, \bibinfo{pages}{3352} (\bibinfo{year}{1995}).

\bibitem[{\citenamefont{Fortagh and Zimmermann}(2007)}]{Zimmermann2007}
\bibinfo{author}{\bibfnamefont{J.}~\bibnamefont{Fortagh}} \bibnamefont{and}
  \bibinfo{author}{\bibfnamefont{C.}~\bibnamefont{Zimmermann}},
  \bibinfo{journal}{Rev. Mod. Phys.} \textbf{\bibinfo{volume}{79}},
  \bibinfo{pages}{235} (\bibinfo{year}{2007}).

\bibitem[{\citenamefont{Trebbia et~al.}(2007)\citenamefont{Trebbia,
  Garrido~Alzar, Cornelussen, Westbrook, and Bouchoule}}]{Bouchoule2007a}
\bibinfo{author}{\bibfnamefont{J.}~\bibnamefont{Trebbia}},
  \bibinfo{author}{\bibfnamefont{C.~L.} \bibnamefont{Garrido~Alzar}},
  \bibinfo{author}{\bibfnamefont{R.}~\bibnamefont{Cornelussen}},
  \bibinfo{author}{\bibfnamefont{C.~I.} \bibnamefont{Westbrook}},
  \bibnamefont{and}
  \bibinfo{author}{\bibfnamefont{I.}~\bibnamefont{Bouchoule}},
  \bibinfo{journal}{Phys. Rev. Lett.} \textbf{\bibinfo{volume}{98}},
  \bibinfo{pages}{263201} (\bibinfo{year}{2007}).

\bibitem[{\citenamefont{Bouchoule et~al.}(2008)\citenamefont{Bouchoule,
  Trebbia, and Garrido~Alzar}}]{Bouchoule2007}
\bibinfo{author}{\bibfnamefont{I.}~\bibnamefont{Bouchoule}},
  \bibinfo{author}{\bibfnamefont{J.-B.} \bibnamefont{Trebbia}},
  \bibnamefont{and} \bibinfo{author}{\bibfnamefont{C.~L.}
  \bibnamefont{Garrido~Alzar}}, \bibinfo{journal}{Phys. Rev. A}
  \textbf{\bibinfo{volume}{77}}, \bibinfo{pages}{023624}
  (\bibinfo{year}{2008}).

\bibitem[{\citenamefont{Murch et~al.}(2006)\citenamefont{Murch, Moore, Gupta,
  and Stamper-Kurn}}]{Stamper-Kurn2006}
\bibinfo{author}{\bibfnamefont{K.~W.} \bibnamefont{Murch}},
  \bibinfo{author}{\bibfnamefont{K.~L.} \bibnamefont{Moore}},
  \bibinfo{author}{\bibfnamefont{S.}~\bibnamefont{Gupta}}, \bibnamefont{and}
  \bibinfo{author}{\bibfnamefont{D.~M.} \bibnamefont{Stamper-Kurn}},
  \bibinfo{journal}{Phys. Rev. Lett.} \textbf{\bibinfo{volume}{96}},
  \bibinfo{pages}{013202} (\bibinfo{year}{2006}).

\bibitem[{\citenamefont{Sackett}(2009)}]{Sackett2009}
\bibinfo{author}{\bibfnamefont{C.~A.} \bibnamefont{Sackett}},
  \bibinfo{journal}{Private Communication}  (\bibinfo{year}{2009}).

\end{thebibliography}

\end{document}